\pdfoutput=1
\documentclass[twocolumn]{aastex61}

\received{}
\revised{}
\accepted{}
\shorttitle{Verification of a stabilized frequency reference for the SKA}
\shortauthors{D. R. Gozzard et al.}
\usepackage{amsmath}

\begin{document}

\title{Astronomical verification of a stabilized frequency reference transfer system for the Square Kilometre Array}

\correspondingauthor{David R. Gozzard}
\email{david.gozzard@research.uwa.edu.au}

\author[0000-0001-8828-6469]{David R. Gozzard}
\affiliation{School of Physics and Astrophysics, University of Western Australia, Perth, WA 6009, Australia}
\affiliation{International Centre for Radio Astronomy Research, University of Western Australia, Perth, WA 6009, Australia}

\author[0000-0001-9928-1364]{Sascha W. Schediwy}
\affiliation{International Centre for Radio Astronomy Research, University of Western Australia, Perth, WA 6009, Australia}
\affiliation{School of Physics and Astrophysics, University of Western Australia, Perth, WA 6009, Australia}

\author[0000-0003-0392-3604]{Richard Dodson}
\affiliation{International Centre for Radio Astronomy Research, University of Western Australia, Perth, WA 6009, Australia}

\author[0000-0003-4871-9535]{Mar\'{\i}a J. {Rioja}}
\affiliation{International Centre for Radio Astronomy Research, University of Western Australia, Perth, WA 6009, Australia}
\affiliation{CSIRO Astronomy and Space Science, 26 Dick Perry Avenue, Kensington WA 6151, Australia}
\affiliation{Observatorio Astron\'omico Nacional (IGN), Alfonso XII, 3 y 5, 28014 Madrid, Spain}

\author{Mike Hill}
\affiliation{CSIRO Australia Telescope National Facility, Paul Wild Observatory, Narrabri NSW 2390, Australia}

\author{Brett Lennon}
\affiliation{CSIRO Astronomy and Space Science, Epping NSW 1710, Australia}

\author{Jock McFee}
\affiliation{CSIRO Australia Telescope National Facility, Paul Wild Observatory, Narrabri NSW 2390, Australia}

\author{Peter Mirtschin}
\affiliation{CSIRO Australia Telescope National Facility, Paul Wild Observatory, Narrabri NSW 2390, Australia}

\author{Jamie Stevens}
\affiliation{CSIRO Australia Telescope National Facility, Paul Wild Observatory, Narrabri NSW 2390, Australia}

\author[0000-0002-6780-1406]{Keith Grainge}
\affiliation{Jodrell Bank Centre for Astrophysics, Alan Turing Building, School of Physics \& Astronomy, The University of Manchester, Oxford Road, Manchester M13 9PL, UK}

\begin{abstract}

In order to meet its cutting-edge scientific objectives, the Square Kilometre Array (SKA) telescope requires high-precision frequency references to be distributed to each of its antennas. The frequency references are distributed via fiber-optic links and must be actively stabilized to compensate for phase-noise imposed on the signals by environmental perturbations on the links. SKA engineering requirements demand that any proposed frequency reference distribution system be proved in ``astronomical verification'' tests. We present results of the astronomical verification of a stabilized frequency reference transfer system proposed for SKA-mid. The dual-receiver architecture of the Australia Telescope Compact Array was exploited to subtract the phase-noise of the sky signal from the data, allowing the phase-noise of observations performed using a standard frequency reference, as well as the stabilized frequency reference transfer system transmitting over 77~km of fiber-optic cable, to be directly compared. Results are presented for the fractional frequency stability and phase-drift of the stabilized frequency reference transfer system for celestial calibrator observations at 5~GHz and 25~GHz. These observations plus additional laboratory results for the transferred signal stability over a 166~km metropolitan fiber-optic link are used to show that the stabilized transfer system under test exceeds all SKA phase-stability requirements under a broad range of observing conditions. Furthermore, we have shown that alternative reference dissemination systems that use multiple synthesizers to supply reference signals to sub-sections of an array may limit the imaging capability of the telescope.

\end{abstract}


\section{Introduction}

The Square Kilometre Array (SKA) project \citep{dewdneybaselinev2} is an international initiative
   to build the largest and most capable radio telescope ever constructed. 
   Construction of the SKA has been divided into phases, with the first phase
   (SKA1) accounting for the first 10\% of the telescope's receiving capacity. 
   During SKA1, a low-frequency aperture array (SKA1-low) will be constructed 
   in Western Australia, while a dish-array of 197 antennas 
   (SKA1-mid), incorporating the 64 dishes of MeerKAT, will be constructed in South Africa. 
   
   \begin{figure*}
   \centering
   \includegraphics[width=\hsize]{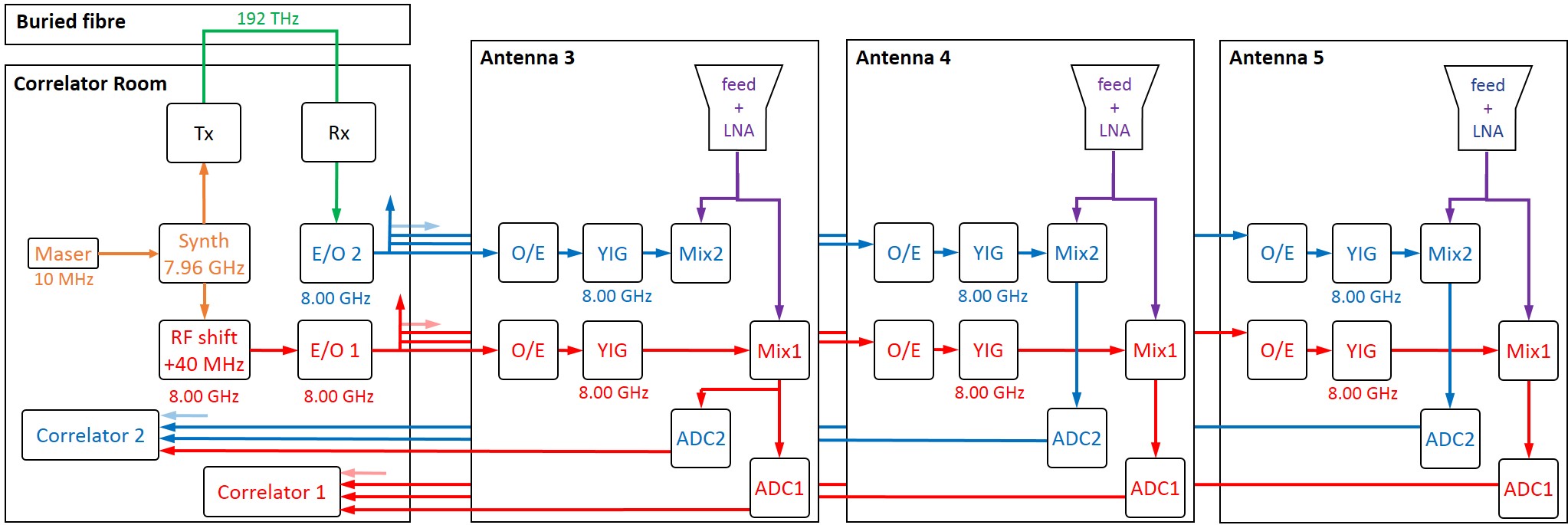}
      \caption{Schematic of the reference frequency distribution setup for astronomical verification at the ATCA. Three antennas are shown for reference. Antenna 3 shows the modification made to antennas 1 and 3 to enable comparative measurements of the two frequency references to be made. Baselines with antennas 1 or 3 using correlator 2 are `mixed' reference baselines. Tx, stabilized frequency transfer system transmitter unit; Rx, stabilized frequency transfer system receiver unit; RF shift, IQ-mixer producing carrier-supressed single-sideband modulation; E/O, electronic-to-optical transmitter; O/E, optical-to-electronic receiver; YIG, yttrium-iron-garnet clean-up oscillator; LNA, low-noise amplifier; Mix, microwave signal mixer; and ADC, analogue-to-digital converter.
              }
         \label{ATCAschem}
   \end{figure*}
   
   Radio telescope arrays such as the SKA require phase-coherent frequency 
   reference signals to be transmitted to each of the antennas in the array. 
   In the case of the SKA, the frequency references will be generated at a 
   central site and transmitted to the antenna sites via fiber-optic cables. 
   Environmental influences affect the optical path length of the fiber and 
   act to degrade the phase stability of the frequency references received 
   at the antennas, which has the ultimate effect of reducing the fidelity and dynamic range of the data \citep{1580149}. To improve the phase-coherence of the array,
   the SKA will employ stabilized frequency transfer systems to suppress the
   fiber-optic link noise \citep{grainge2017square}. Reference frequency stabilization systems have been
   successfully used on other radio telescope arrays such as the Very Large Array \citep{thompson1980very}, e-Merlin in the UK \citep{garrington2004merlin}, the Australia Telescope Compact Array (ATCA) \citep{hancock2011australia}
   and the Atacama Large Millimetre Array (ALMA) \citep{1580149,1396874}. The existing SKA precursor telescopes including the Murchison Widefield Array \citep{tingay2013murchison,lonsdale2009murchison}, the Australian SKA Pathfinder \citep{beresford2008askap,hotan2014australian}, and MeerKAT \citep{julie2017measuring} currently employ passive frequency reference dissemination systems that provide adequate phase stability over the relatively short transmission distances required by these telescopes.
   
   There are three SKA systems requirements that apply to the phase-noise performance of the frequency transfer system \citep{SKA1reqsV6}. The frequency transfer system must provide a maximum coherence loss of 2\% over the correlator integration time (1~s) and over the the in-beam calibration time (60~s) for observing frequencies up to 13.8~GHz. Of this 2\% coherence loss budget, 1.9\% is allocated to the stabilized frequency reference transfer system.
   In addition, the frequency reference must have a phase drift of less than 1 rad over a 10-minute period, to ensure that there is no ambiguity in the phase solution during array calibration measurements either side of an observation of up to 10 minutes.
   
   Researchers at the University of Western Australia (UWA) have led the development of a stabilized frequency transfer system proposed for SKA1-mid \citep{schediwy2017stabilized}. This has been tested extensively using standard metrology techniques in a laboratory setting, with signals transmitted over metropolitan fiber links. Laboratory and field testing of an alternative stabilized frequency transfer system for the SKA has also been reported previously \citep{gao2016field,wang2015square} and other research groups have developed similar systems with a view to supporting the SKA and other VLBI applications \citep{he2013stable,baldwin2016dissemination}. However, SKA technology down-select requirements demand a process of ``astronomical verification'' in which the proposed frequency transfer system is shown to meet the stability requirements under observing conditions on an operational radio telescope. In this paper, we present results of astronomical verification of the UWA SKA1-mid stabilized frequency transfer prototype and show that the system exceeds SKA phase stability requirements under a wide range of observing conditions.

%__________________________________________________________________

\section{Experimental design} \label{sec:design}

\subsection{Extracting the reference signal phase stability}

   The astronomical verification observations were performed using the ATCA. The ATCA's dual-receiver chain architecture permitted analyses that would not have been possible to conduct with single-receiver telescope arrays. As shown in Figure~\ref{ATCAschem}, immediately following the feed horn and the first low-noise amplifier (LNA), the sky signal, for both polarizations, is split into two separate, but functionally identical, receiver chains (referred to here as 'chain-1' and 'chain-2'). The sky signals are then down-converted by mixing them with separate frequency reference signals, Frequency Reference 1 and Frequency Reference 2. These are normally supplied by two separate microwave frequency synthesizers located in the observatory's correlator room. The two frequency references are transmitted up to 4.5~km to the antennas through parallel fiber cores in buried conduit using a pair of electronic-to-optical transmitters (E/Os). The transmitted references are detected at the antennas by optical-to-electronic (O/E) receivers (two in each antenna) that feed the signals to a pair of yttrium-iron-garnet (YIG) oscillators operating in a phase-locked-loop. The YIG oscillators act as clean-up oscillators to suppress high-frequency phase noise. The down-converted sky signals are digitized by analogue-to-digital converters (ADCs) that feed to two separate correlators, one for each receiver chain, in the correlator room.
   
   The advantage of this configuration for astronomical verification tests is that Frequency Reference 1 can be supplied by the ATCA's conventional reference signal distribution system, while Frequency Reference 2 is supplied by the stabilized frequency reference transfer system under test (as shown in Figure~\ref{ATCAschem}). Since the two receiver chains in the antennas detect the same sky signal, any differences in the phase solution of these signals between the two receiver chains for a given telescope baseline can be attributed to non-common phase noise produced in the separate frequency reference dissemination systems. For example, the phase solution for the chain-1 baseline between two antennas i and j ($\phi_{ij.1}(t)$) is
   
   \begin{equation}
   \begin{split}
   \phi_{ij.1}(t) ={} & (\phi_{R1}(t)+\phi_{E1.i}(t)+\phi_{Si}(t)) \\
   &-(\phi_{R1}(t)+\phi_{E1.j}(t)+\phi_{Sj}(t)),
   \end{split}
   \end{equation}
   
   where $\phi_{R}(t)$ is the phase of the Frequency Reference as a function of time (t) and $\phi_{S}(t)$ is the phase of the sky signal at a particular antenna receiver. The term $\phi_{E}(t)$ represents the phase-noise contributed by components in the ATCA reference frequency distribution system, including the E/O/E systems, differential fiber-link noise, and noise from the YIG oscillator. The corresponding phase solution for the chain-2 baseline between antennas i and j ($\phi_{ij.2}(t)$) is
   
   \begin{equation}
   \begin{split}
   \phi_{ij.2}(t) ={} & (\phi_{R2}(t)+\phi_{E2.i}(t)+\phi_{Si}(t)) \\
   & - (\phi_{R2}(t)+\phi_{E2.j}(t)+\phi_{Sj}(t)).
   \end{split}
   \end{equation}
   
   In the configuration depicted in Figure~\ref{ATCAschem}, any additional phase-noise contributed by the stabilized frequency transfer system (Frequency Reference 2) is common to all of the down-converted signals from the chain-2 mixer outputs. Therefore, the chain-2 baseline phase solutions do not exhibit additional phase-noise compared with the simultaneous phase solution for the same physical baseline using chain-1 (as shown by equations 1 and 2). 
   
   To remedy this, the outputs of the chain-1 mixers in antennas 1 and 3 were split using microwave power splitters in the antennas, and fed to both the chain-1 and chain-2 ADCs. This was done for both the A- and B-polarisations (for clarity, only one polarization is shown in Figure~\ref{ATCAschem}). As a result, the outputs of the chain-2 ADCs on antennas 1 and 3 did not incorporate the phase-noise contributed by the stabilized frequency transfer system and the chain-2 E/O system. The phase solution $\phi_{ij.2}(t)$ (where antenna \textit{i} is either antenna 1 or 3) then becomes:
   
   \begin{equation}
   \begin{split}
   \phi_{ij.2}(t) ={} & (\phi_{R1}(t)+\phi_{E1.i}(t)+\phi_{Si}(t)) \\
   &-(\phi_{R2}(t)+\phi_{E2.j}(t)+\phi_{Sj}(t)).
   \end{split}
   \end{equation}
   
   By subtracting the chain-1 phase solution from the chain-2 phase solution, for baselines that included antennas 1 or 3, the differential phase-noise between the two frequency references could be measured. Hereafter, chain-2 baselines that incorporate antenna 1 or antenna 3 will be referred to as `mixed' baselines (because they operate using a `mix' of two different frequency references). Baselines operating in the conventional manner will be referred to as `unmixed' baselines. The resulting phase difference between the phase solutions for the unmixed and mixed baselines is then:
   
   \begin{equation}
   \begin{split}
   \phi_{Diff}(t) &= \phi(t)_{ij.1}-\phi_{ij.2}(t) \\
   &=\phi_{R2}(t)-\phi_{R1}(t)+\phi_{E2.j}(t)-\phi_{E1.j}(t).
   \end{split}
   \end{equation}
   
   This means that, not only can a direct comparison be made of the array performance under the two different frequency transfer systems, but the phase of the sky signal has been subtracted out, and so the relative stability of the frequency references (with non-common phase-noise contributions from some telescope systems) can be measured directly.
   
   To avoid the contribution of the relative phase drift between two synthesizers, a single microwave synthesizer (Agilent E8257D) was used to supply both Frequency Reference 1 and the UWA stabilized frequency transfer prototype (which supplies Frequency Reference 2). So, the phase of Frequency Reference 2 is
   
   \begin{equation}
   \phi_{R2}(t)=\phi_{R1}(t)+\phi_{UWA}(t),
   \end{equation}
   
   where $\phi_{UWA}(t)$ is the phase of the UWA stabilized frequency transfer prototype. The phase difference becomes:
   
   \begin{equation}
   \phi_{Diff}(t)=\phi_{UWA}(t)+\phi_{E2.j}(t)-\phi_{E1.j}(t).
   \end{equation}
   
   Therefore the phase stability of the UWA stabilized frequency transfer prototype can be measured, however, with some contamination ($\phi_{E2.j}(t)-\phi_{E1.j}(t)$) from the standard ATCA frequency reference distribution systems. Because the ATCA receiver chains and frequency reference distribution systems use identical components, and the fiber links from the correlator room to the antennas run through the same buried cables, we assume that the contributions of these components to the phase differences are small and the differences are dominated by phase-noise contributed by the stabilized frequency transfer system.
   
   \subsection{Experimental setup}
   
   The astronomical source used to make the measurement observations was a calibrator labelled 1057-797, at a declination of close to $-80$\degr, which is isolated in the sky and always above the ATCA horizon, enabling long uninterrupted array phase measurements. The frequency reference signal was transmitted at 8.00~GHz because this is the nominal frequency of UWA's SKA1-mid stabilized frequency transfer system. Given the specific design of the ATCA's receiver architecture \citep{ATCAguide}, this enabled observations across two separate frequency bands, with centre observing frequencies of 4.96~GHz and 25.44~GHz and a bandwidth of 2048~MHz. The lower observing frequency is representative of SKA1-mid's operational observing range of 0.35~GHz to 13.8~GHz \citep{dewdneybaselinev2}, while the higher observing frequency demonstrates the performance of the stabilized frequency transfer system under more demanding observing conditions than SKA1-mid will encounter.
   
   The transmitter unit (Tx) of the UWA SKA1-mid stabilized frequency transfer prototype was installed in the ATCA correlator room. In order for Frequency Reference~1 and Frequency Reference~2 to both supply 8.00~GHz while using only one microwave synthesizer, the synthesizer was set to supply 7.96~GHz which the servo system of the stabilized frequency transfer prototype shifted up to 8.00~GHz to supply Frequency Reference 2. An electronic IQ-mixer (Marki IQ0741LXP), set to produce carrier-supressed single-sideband modulation at 40~MHz, was used to shift the synthesizer signal to 8.00~GHz to supply Frequency Reference 1. Laboratory measurements of the intrinsic stability of the IQ-mixer frequency shift show that it is better than the stability of the stabilized frequency transfer system by an order of magnitude, so did not make a significant contribution to the measured phase-noise.
   
   The Tx and the receiver unit (Rx) were co-located in the correlator room as shown in Figure~\ref{ATCAschem}. The signal transmitted by the stabilized frequency transfer prototype was sent over 52~km of buried fiber-optic cable and a further 25~km spool of fiber in the correlator room, producing a total link length of 77~km. The 52~km fiber link consisted of two 26~km fiber cores that ran off-site, under and along local roadways and a bridge (through a variety of moderate-to-high shrink-swell soil types), to a telecommunications controlled environment vault at Springbrook Creek (located along the Newell Highway, approximately 17.5~km south-east of the observatory and 16.3~km south-west of Narrabri). A short fiber patch was used to connect the two cores in the vault, creating a 52~km loop. A bi-directional erbium-doped fiber amplifier (IDIL Fibres Optiques) was required to compensate for the optical power loss of the 77~km link.
   
   \begin{table*}
   \caption{Summary of astronomical verification observing runs.}             
   \label{table:1}      
   \centering          
   \begin{tabular}{c c c c c c }
   \hline\hline 
   Date & Centre observing & Bandwidth & Available antennas & Logging & Details\\
      & frequency (GHz) & (MHz) &   & rate (s) &   \\
   \hline                    
   2016-05-26 & 4.96 & 2048 & 01, 02, 03, 04, 05, 06 & 12 & 12 hr overnight run, dual synthesizers\\
   2016-06-20 & 4.96 & 2048 & 01, 03, 04, 06    & 5 & 13 hour overnight run\\
   2016-06-21 & 4.96 & 2048 & 01, 03, 04, 06     & 5 & 12 hr overnight run\\
   2016-06-22 & 4.96 & 2048 & 01, 03, 04, 05, 06    & 1 & 2 hr daytime run \\
   2016-06-22 & 25.44 & 2048 & 01, 03, 04, 05, 06    & 5 & 15 hr overnight run\\
   \hline                  
   \end{tabular}
   \label{table1}
   \end{table*}
   
   In addition, measurements were made to assess the effect of phase drift between two microwave synthesizers. One supplied the standard ATCA frequency distribution system with 8.00~GHz and the second supplied the stabilized frequency distribution system with 7.96~GHz. Both synthesizers were referenced to the observatory's hydrogen masers.
   
   Observations were performed over several days in May and June 2016 during scheduled ATCA maintenance periods. Table~\ref{table1} summarizes the measurements that will be discussed in this paper. Due to maintenance requirements, the number of antennas available at different times varied.
   
   The logging rate was limited by the data volume limitation of the correlator and it was not possible to perform overnight observations with a logging period shorter than five seconds. A shorter observation of only two hours was performed with a logging period of one second in order to obtain array phase stability measurements at shorter integration times.
   
   \begin{figure*}
   \centering
   \includegraphics[width=\hsize]{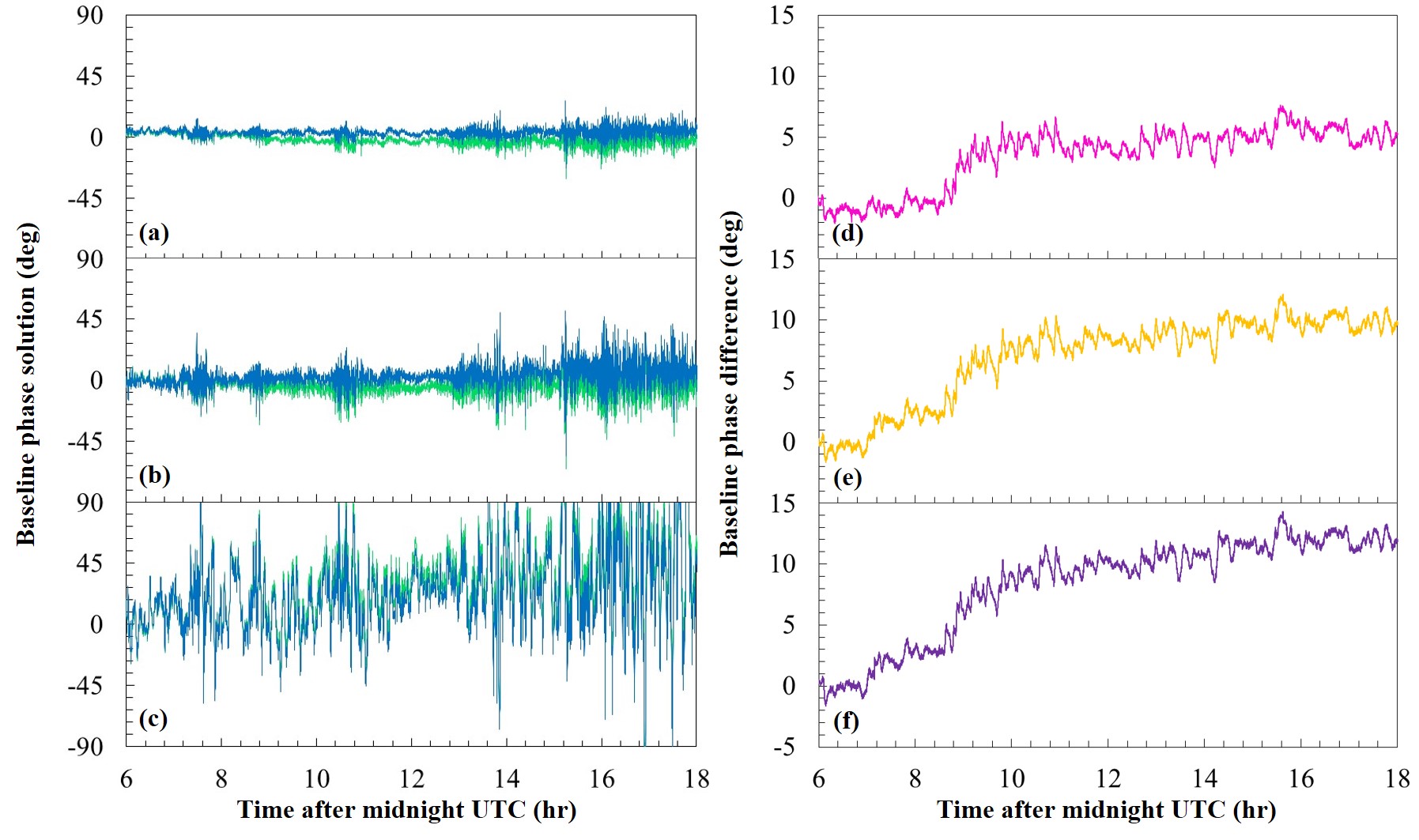}
      \caption{Phase solutions (a-c) and corresponding phase differences (d-f) for a selection of three baselines of different lengths (a,~d~=~46~m; b,~e~=~152~m; c,~f~=~4439~m) for 4.96~GHz observations from 21st June 2016. Blue -- unmixed baselines; green -- mixed baselines; pink (d), orange (e), and purple (f) -- phase differences for baselines (a), (b) and, (c) respectively.
              }
         \label{J21Phases}
   \end{figure*}

%
%______________________________________________________________

\section{Results} \label{sec:results}

Phase solutions and phase differences were produced for all mixed and unmixed baselines for each of the measurements summarized in Table~\ref{table1}. An example of the phase solutions and phase differences for a single polarization for three baselines of different lengths from observations conducted on 21st June 2016 is shown in Figure~\ref{J21Phases}. The blue traces represent the unmixed baselines (using the conventional ATCA reference distribution) while the green traces represent the same physical baselines using the mixed references. The corresponding phase differences are shown to the right of the figure.

At the commencement of the observations, the delay errors for each antenna were measured by the correlator while observing the calibrator source, and corrected for (residual delay errors were typically less than 0.1 nanoseconds). We also added phase offsets to produce a zero phase for each baseline at the beginning of the observation. No further delay or phase calibration was made while the array was observing.

\begin{figure*}
   \centering
   \includegraphics[width=15cm]{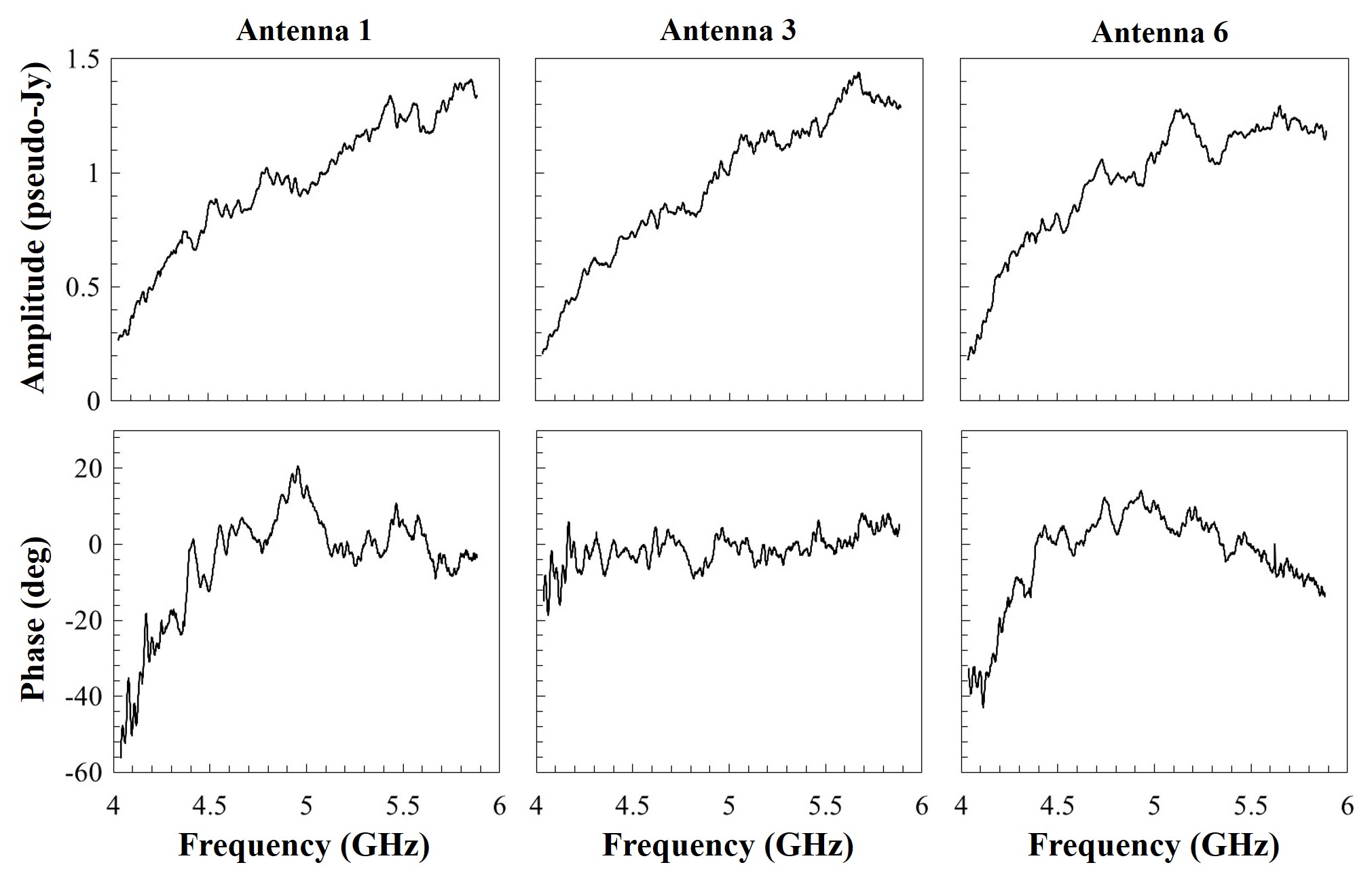}
      \caption{Example amplitude and phase bandpass solutions from 4.96~GHz overnight observation from 21st June 2016. The bandpass solutions for the X-polarization of three antennas (antennas 1, 3, and 6) are shown.
              }
         \label{AmpPhaseBandPass}
   \end{figure*}

All post-observation data reduction was performed with the Miriad software package \citep{sault1995retrospective}. First, we performed a bandpass calibration with the Miriad task mfcal, while solving for time-varying gains every 30 seconds. The resultant bandpass solutions (examples of which are shown in Figure~\ref{AmpPhaseBandPass}) were then subtracted from the entire dataset (we assume that they stay constant with time). We then used the task gpcal to solve for time-variable complex gains every cycle, which is possible due to the brightness of the calibrator source. This was used to produce the phase solutions (like those shown in Figure~\ref{J21Phases}) that reveal how the measured phases changed over the duration of the observation.

\begin{figure*}
   \centering
   \includegraphics[width=\hsize]{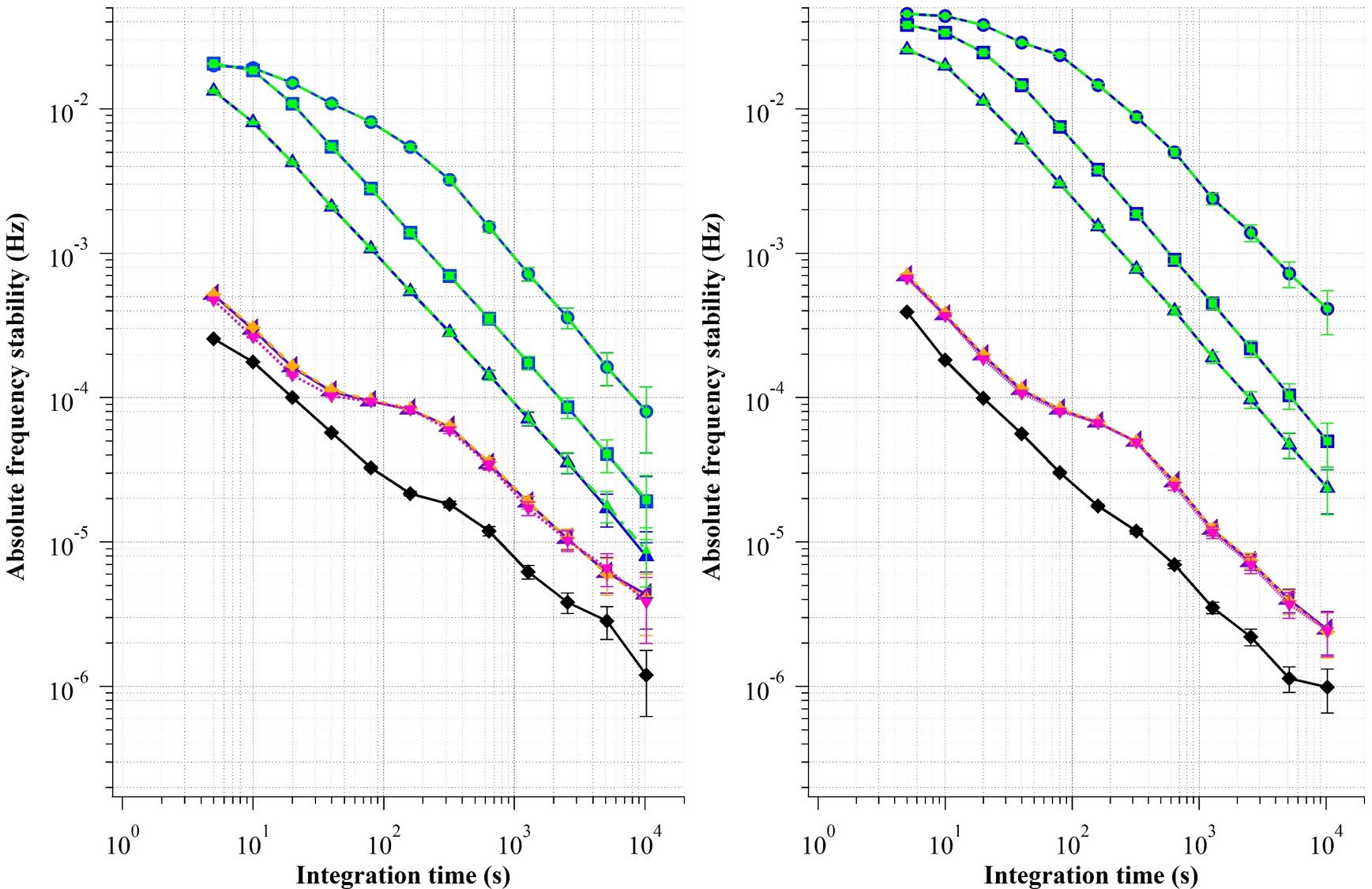}
      \caption{Absolute frequency stability of phase solutions and phase differences for (i) 4.96 GHz overnight observation from 21st June 2016 and (ii) 25.44 GHz overnight observation from 22nd June 2016. Phase solutions: blue triangles, solid line -- unmixed 46 m baseline; green triangles, dashed line -- mixed 46~m baseline; blue squares, solid line -- unmixed 152~m baseline; green squares, dashed line -- mixed 152 m baseline; blue circles, solid line -- unmixed 4439~m baseline; and green circles, dashed line -- mixed 4439~m baseline. Phase differences: pink triangles, dotted line -- 46~m baseline; orange stars, dashed line -- 152~m baseline; purple triangles, solid line -- 4439~m baseline; and black diamonds, solid line -- 107~m baseline between antennas 1 and 3 using only the standard ATCA reference system.
              }
         \label{AbsFreqStab1}
   \end{figure*}

The phase solutions and phase differences were processed to produce plots of absolute frequency stability. Figure~\ref{AbsFreqStab1} shows the absolute stabilities for the three baselines used as examples in Figure~\ref{J21Phases} as well as the phase difference stability of an unmixed (conventionally referenced) baseline for comparison. Absolute frequency stability values are obtained by multiplying the Allan deviation computed for the signal by the frequency of the signal. Values in terms of absolute frequency stability are necessary to directly compare the stability of signals of different frequencies and allows the noise contributions of different parts of the interferometer and frequency reference systems to be assessed more easily. Figure~\ref{AbsFreqStab1} shows the absolute stability plots for the traces shown in Figure~\ref{J21Phases}, as well as for the 25 GHz measurements from 22nd June 2016 (using the same example baselines).

\begin{figure}
   \centering
   \includegraphics[width=\hsize]{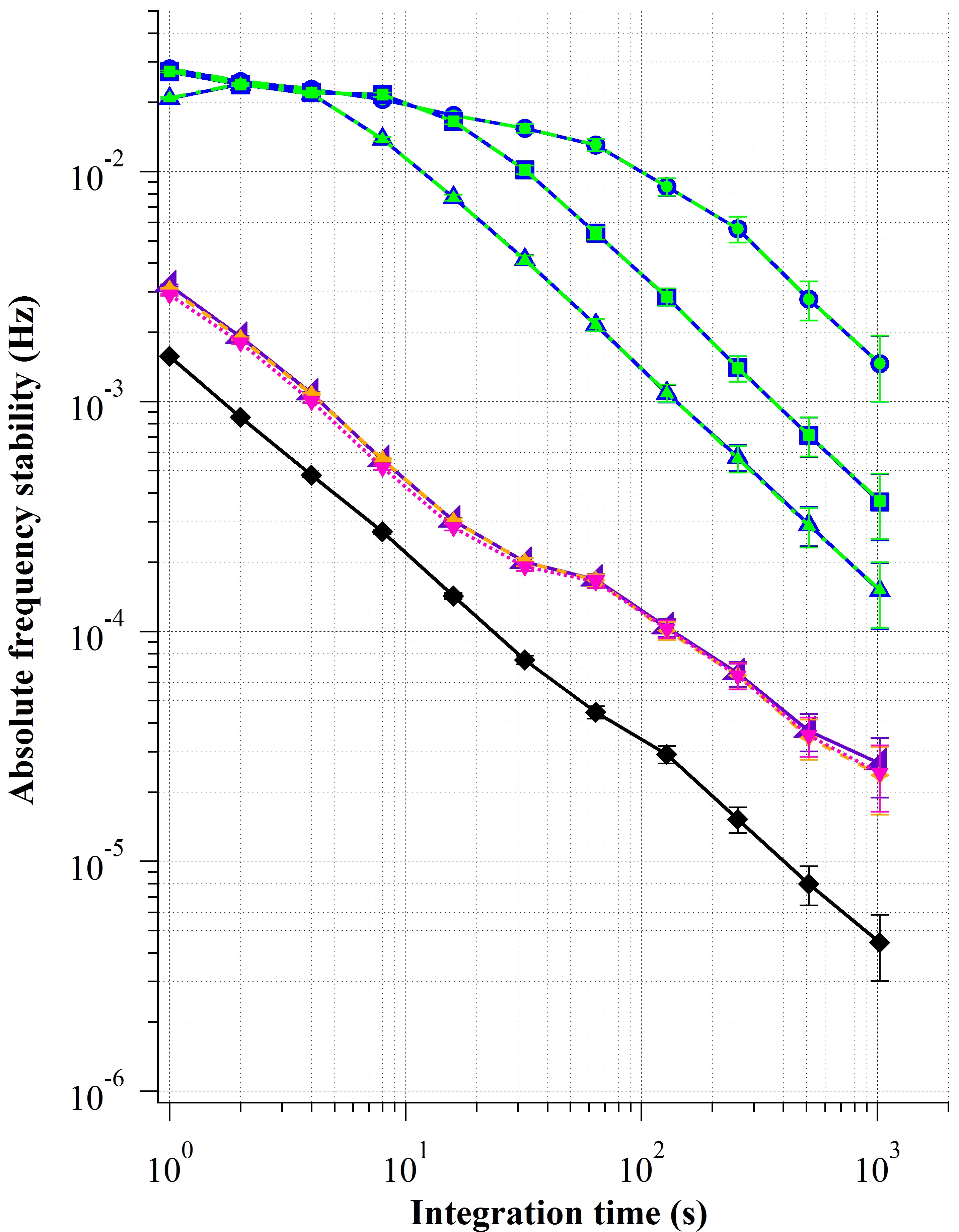}
      \caption{Absolute frequency stability of phase solutions and phase differences for one-second logged 4.96 GHz daytime observation from 22nd June 2016. Phase solutions: blue triangles, solid line -- unmixed 46 m baseline; green triangles, dashed line -- mixed 46~m baseline; blue squares, solid line -- unmixed 152~m baseline; green squares, dashed line -- mixed 152 m baseline; blue circles, solid line -- unmixed 4439~m baseline; and green circles, dashed line -- mixed 4439~m baseline. Phase differences: pink triangles, dotted line -- 46~m baseline; orange stars, dashed line -- 152~m baseline; purple triangles, solid line -- 4439~m baseline; and black diamonds, solid line -- 107~m baseline between antennas 1 and 3 using only the standard ATCA reference system.
              }
         \label{1sFreqStab}
   \end{figure}

The process was repeated for the one-second logged daytime run from 22nd June 2016 and the resulting absolute frequency stabilities are shown in Figure~\ref{1sFreqStab}. The one-second logging period of this measurement allows the stability values to be calculated at integration times down to one second but, due to the relatively short time span of the observations, stability data are limited to integration times shorter than 1,024~s.

\begin{figure}
   \centering
   \includegraphics[width=\hsize]{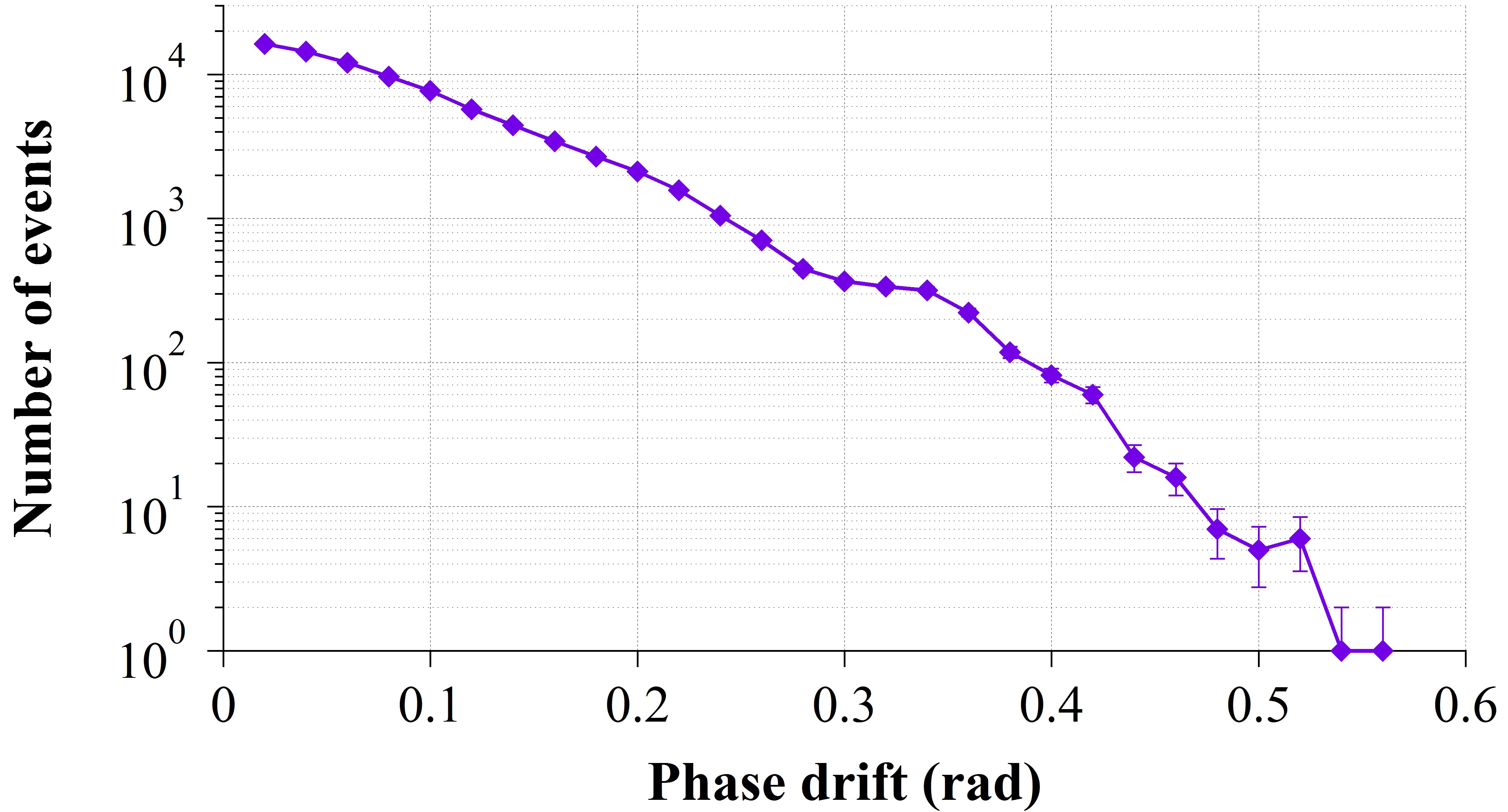}
      \caption{Total number of phase drifts of a particular magnitude over 10 minutes for all observations with the stabilized frequency transfer system.
              }
         \label{PhaseDrifts}
   \end{figure}

To assess the performance of the system with respect to the 10 minute phase drift requirement, the phase drifts between the mixed and un-mixed baselines for each 10 minute interval in all 42 hours of stabilized observations (Table~\ref{table1}, 20th - 22nd June 2016) were measured. The compiled phase drift values are shown in Figure~\ref{PhaseDrifts}. Within one standard deviation, the phase drifts are less than 0.08~rad, and no 10-minute period phase drift exceeded a value of 0.56~rad.

\begin{figure}
   \centering
   \includegraphics[width=\hsize]{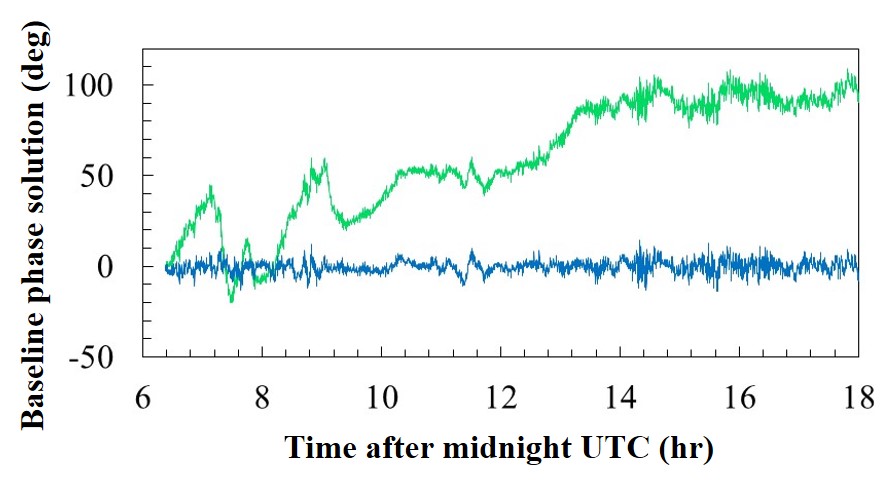}
      \caption{Phase solutions for a single polarization for a 76 m baseline from the dual-synthesizer observation from 26th May 2016. Blue -- unmixed baseline, and green -- mixed baseline.
              }
         \label{DualSynthPhases}
   \end{figure}

\begin{figure}
   \centering
   \includegraphics[width=\hsize]{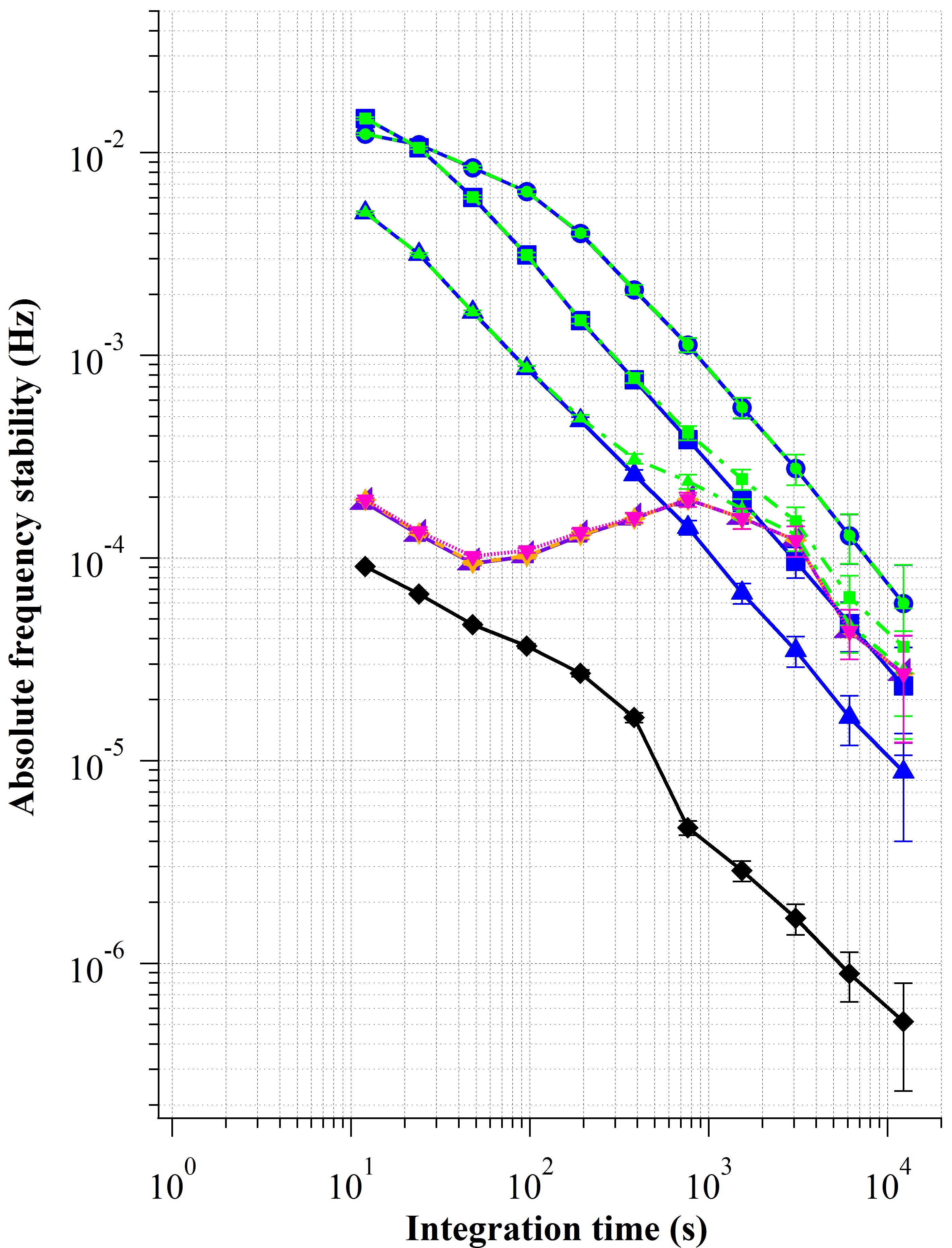}
      \caption{Absolute frequency stability of phase solutions and phase differences for 4.96 GHz overnight dual-synthesizer observation from 26th May 2016. Phase solutions: blue triangles, solid line -- unmixed 76~m baseline; green triangles, dashed line -- mixed 76~m baseline; blue squares, solid line -- unmixed 138~m baseline; green squares, dashed line -- mixed 138~m baseline; blue circles, solid line -- unmixed 3505~m baseline; and green circles, dashed line -- mixed 3505~m baseline. Phase differences: pink triangles, dotted line -- 76~m baseline; orange stars, dashed line -- 138~m baseline; purple triangles, solid line -- 3505~m baseline; and black diamonds, solid line -- 413~m baseline between antennas using only the standard ATCA reference system.
              }
         \label{AbsFreqStab2}
   \end{figure}

The synthesizer comparison test (Table~\ref{table1}, 26th May 2016) exhibits large phase drifts between the mixed and unmixed baselines, an example of which is shown in Figure~\ref{DualSynthPhases}. This relative phase drift between the two synthesizers significantly affected the stability of the baseline phase solutions and phase differences. Figure~\ref{AbsFreqStab2} shows the resulting absolute frequency stabilities for three example mixed baselines plus an unmixed baseline for comparison.

%
%______________________________________________________________

\section{Discussion} \label{sec:discuss}

\subsection{Stability analysis}

\begin{figure}
   \centering
   \includegraphics[width=\hsize]{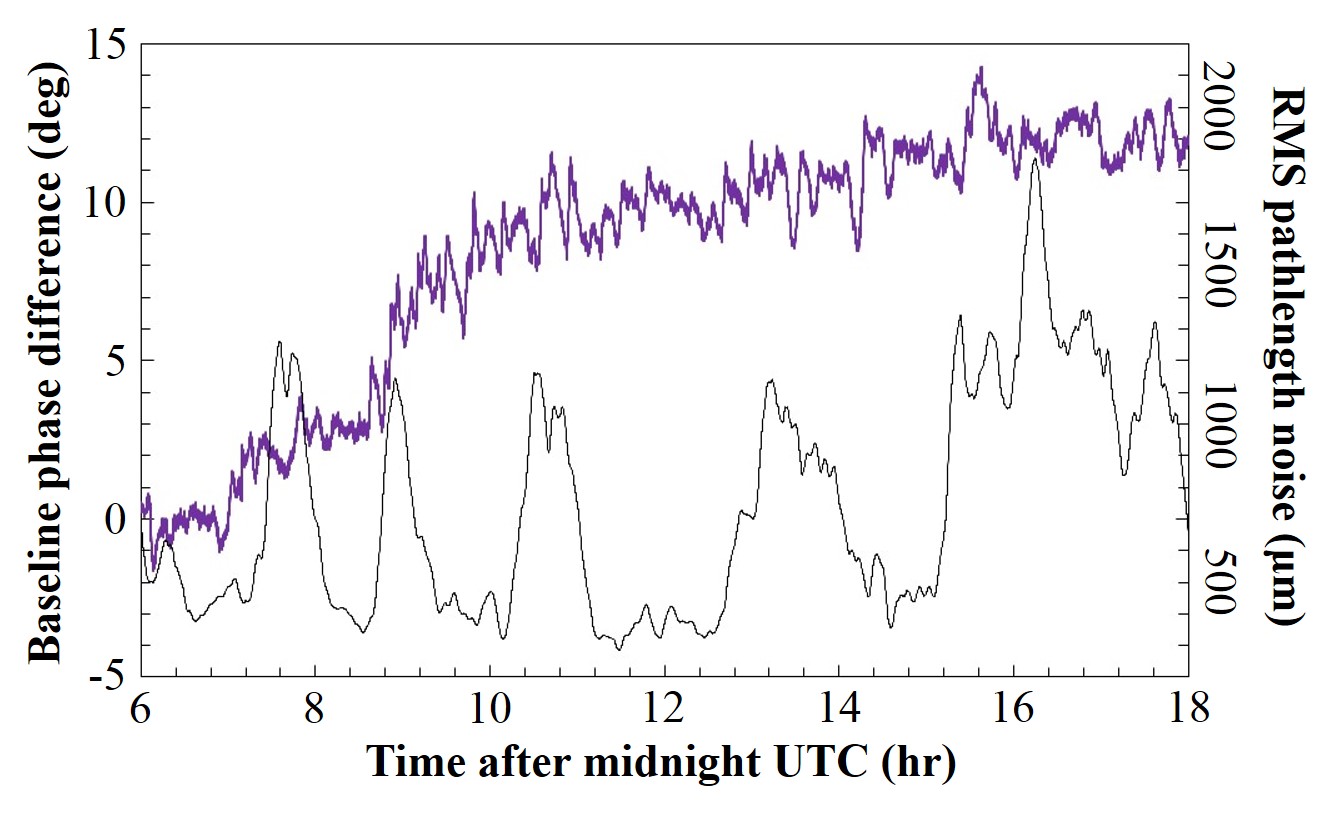}
      \caption{Output of the ATCA seeing monitor (black line) overlaid with the phase differences for the 4439~m baseline from the 4.96~GHz overnight observation from 21st June 2016 (purple trace, copy of Fig.~\ref{J21Phases} panel (f)). Increases in atmospheric path length noise shown by the seeing monitor correspond with increases in the rate of phase drift.
              }
         \label{seeingmon}
   \end{figure}

Each of the example stability plots in Figures~\ref{AbsFreqStab1} and \ref{1sFreqStab} show three traces for the stability of the phase difference between the mixed and unmixed baselines, one each for the short (pink trace), intermediate (orange trace), and long (purple trace) baselines. As expected, the three curves lie on top of each other because they are the result of effectively subtracting-out the phase of the sky signal. The remaining phase-noise is due to the reference distribution systems and non-common elements of the receiver chains. The frequency transfer systems (UWA stabilized transfer system plus ATCA distribution system electronics) are the dominant source of differential noise and, because that noise is common to all chain-2 frequency references, the frequency stability curves for the different baselines are almost identical. Panels (d), (e) and (f) in Figure~\ref{J21Phases} show that the phase differences have some remaining baseline dependence, with the total phase drift increasing with increasing baseline length. This may be due to differential noise on the fiber links between the correlator room and the antennas as well as residual atmospheric effects. Figure~\ref{seeingmon} shows the output of the ATCA seeing monitor \citep{middelberg2006atca,indermuehle2014millimetre} for the time period corresponding to the example observations from 21st June 2016, overlaid with the phase differences for the 4439~m baseline from this observation (Figure~\ref{J21Phases}, panel (f)). The data from the seeing monitor show that the atmospheric path length noise is greater during the periods when there is a noticeable increase in the rate of phase drift.

The difference between the stabilities of the phase differences for the mixed (pink, orange and purple traces in Figures.~\ref{AbsFreqStab1} and \ref{1sFreqStab}) and unmixed baselines (black traces in Figures.~\ref{AbsFreqStab1} and \ref{1sFreqStab}) is due to noise contributed by both the stabilized frequency transfer system and parts of the ATCA frequency distribution system and receiver chains. While the stabilized frequency transfer system is assumed to be the dominant source of the extra phase noise, the following analysis shows that phase noise contributed by the ATCA systems account for a significant fraction of the difference between the mixed and unmixed baseline phase difference stabilities. The mixed baseline phase difference stabilities compared to the integration time ($\tau$) follow a power law with a gradient that is close to $\tau^{-1}$ up to integration times of around 40~s. The phase difference stabilities for the baseline between antennas 1 and 3 (comparing an unmixed baseline with an unmixed baseline, black traces in Figures.~\ref{AbsFreqStab1} and \ref{1sFreqStab}) also exhibit this power law out to longer integration times, indicating that some of this extra noise does not originate with the stabilized frequency transfer system \citep{astrover}. This power law is a signature of white- and/or flicker-phase noise and is most likely due to noise introduced by amplifiers outside of the stabilized frequency transfer system \citep{TMS}. Measurements of the relative phase drift between the two E/O systems, conducted separately to the main observations, indicated that the `bump' feature in the phase difference stabilities between $\tau = $~40~s and $\tau = $~640~s is due to phase-noise in the E/O systems \citep{astrover}. This noise is not normally obvious during conventional observations because there is no cross-correlation between signals from chain-1 and chain-2.

\subsection{Extrapolation to SKA1}

\begin{figure}
   \centering
   \includegraphics[width=\hsize]{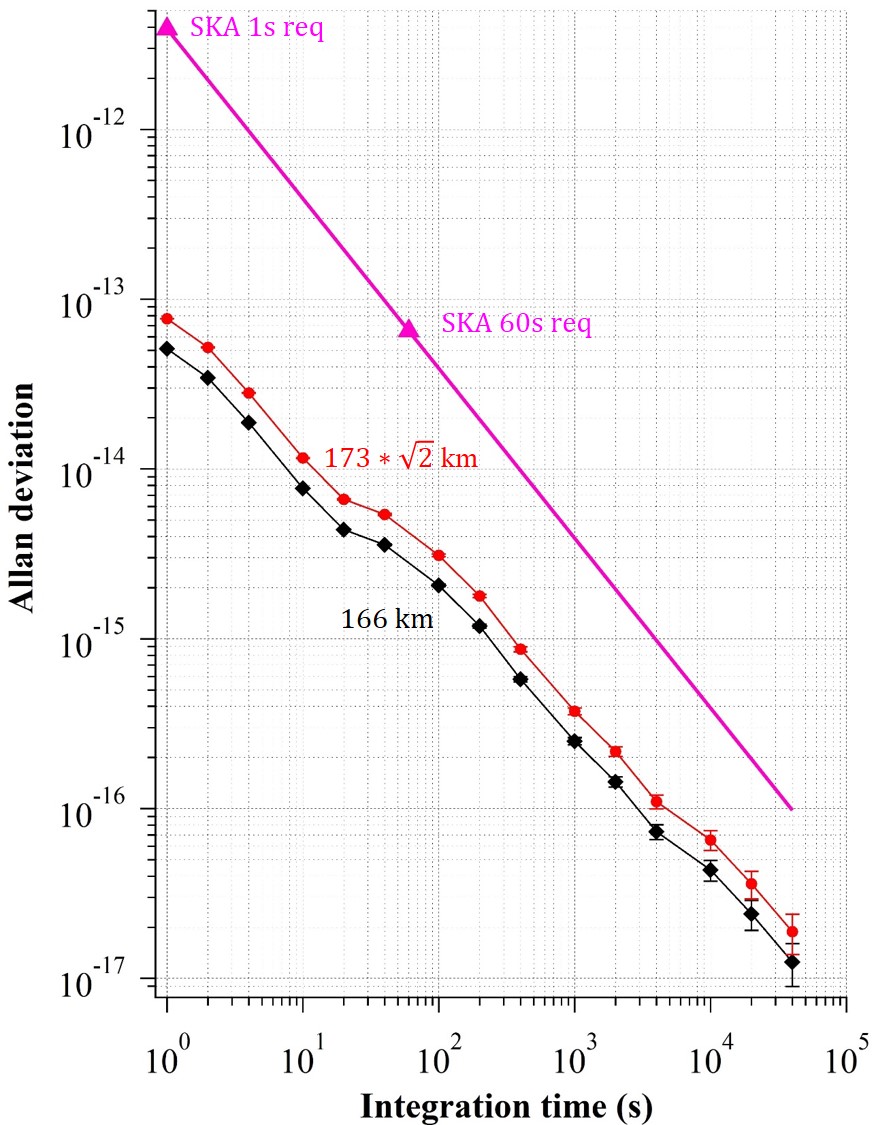}
      \caption{Allan deviation of the stabilized frequency transfer system over 166~km of buried conduit fiber around Perth, Western Australia (black diamonds), and extrapolated to two links of 173~km (red circles). Allan deviation limit corresponding to SKA 1~s and 60~s coherence requirements (assuming white phase noise) represented by pink line.
              }
         \label{173kmExtrap}
   \end{figure}

With the main contributions to the residual differences between the stability of the phase differences for mixed and unmixed baselines accounted for, the measured absolute frequency stabilities are in good agreement with values expected from laboratory measurements of the stabilized frequency transfer system \citep{schediwy2017stabilized}. This provides confidence that standard metrology measurement techniques are adequate to reliably demonstrate the stability of the stabilized frequency transfer system for SKA verification purposes. The stability of the transmission over 166~km of buried conduit fiber-optic cable around Perth, Western Australia (Figure~\ref{173kmExtrap}), has been measured previously using a Microsemi 5125A Phase Noise Test Set. The Allan deviation values for the signal stability over the 166~km link were extrapolated to obtain a prediction for the performance of the stabilized transfer system over a single 173 km span (the longest SKA1 fiber link length). This was achieved by multiplying the measured Allan deviation values by (L2/L1)$^{3/2}$ \citep{williams2008high}, where L1 is the length of the test link (166~km) and L2 is the length of the link for which the prediction is being made (173~km). These predicted Allan deviation values were then multiplied by a factor of $\sqrt{2}$ to give a prediction of the Allan deviation between two stabilization system receiver units at the ends of two separate 173~km links (assuming that deviations in the two signals are uncorrelated).
By assuming that the dominant noise process of the stabilized frequency reference is white phase noise (as indicated by the approximately $\tau^{-1}$ slope of the Allan deviation measurement), an estimate of the coherence loss can be calculated using the process described by \cite{rogers1981coherence} and \cite{TMS}. Using this process, an upper limit was calculated for the permissible Allan deviation of the frequency transfer system corresponding to the 1.9\% coherence loss requirement for a maximum observing frequency of 13.8~GHz. This Allan deviation is $3.91~\times~10^{-12}/\tau$ and is shown in Figure~\ref{173kmExtrap} (pink line). The measured and predicted stability of the stabilized frequency transfer system is well below this limit at all integration times.

Applying the same coherence calculations directly to the measured Allan deviation, the coherence losses resulting from the extrapolated Allan deviation values for a maximum observing frequency of 13.8~GHz are $7.41~\times~10^{-6}$ at one second integration time and $7.96~\times~10^{-5}$ at 60 seconds integration time. These values exceed the 1.9\% coherence loss requirement by a factor 2560 and 239 respectively.

An alternative method of estimating coherence loss from Allan deviation was used by members of the ALMA project \citep{yamada2006phase,kiuchi2005alma2,kawaguchi1983coherence}. Using this method we calculate coherence losses of $4.4~\times~10^{-6}$ at one second integration time and $1.6~\times~10^{-5}$ at 60 seconds integration time. These values exceed the 1.9\% requirement by factors of 4320 and 1190 respectively.

\subsection{Dual-synthesizer test analysis}

The large phase drift in the results from 26th May 2016 (green trace, Figure~\ref{DualSynthPhases}) is dominated by the phase drift between the two synthesizers used as frequency references. In order for a synthesizer to maintain its output frequency relative to a 10 MHz reference, it must change the phase of its output signal as its internal temperature changes in response to variations in ambient temperature. The large drifts were not seen when only one synthesizer was used with the IQ-mixer frequency shift. Figure~\ref{AbsFreqStab2} shows that the phase-noise of the two-synthesizers even dominates the phase-noise of the sky signals caused by the atmosphere on baselines up to around 200~m in length at observing frequencies around 5~GHz. This is not an issue in normal observations because interferometers such as the ATCA do not operate in configurations where multiple synthesizers are used to provide frequency references to different antennas or sub-arrays. However, any system that uses multiple remote synthesizers may encounter such a synthesiser phase drift issue despite the transmission frequency being successfully stabilized. This is a critical consideration for the design of stabilized time and frequency reference systems planned for the SKA.

%______________________________________________________________

\section{Conclusions} \label{sec:conclude}

We have used the dual-receiver architecture of the ATCA to perform astronomical verification tests of a stabilized frequency transfer system proposed for the SKA, as well as to assess the effects of using multiple microwave frequency synthesizers to supply frequency references to different telescope antennas or sub-arrays. The results of these astronomical verification trials show that the stabilized frequency transfer prototype exceeds the SKA level-1 coherence and phase drift requirements under a broad range of observing conditions. Extrapolating the system's performance from laboratory measurements over a link of 166~km to SKA operating conditions predicts that, even for a worst-case scenario, this stabilized frequency transfer system will exceed the 1~s coherence requirement by three orders of magnitude, and the 60~s coherence requirement by two orders of magnitude. Over the 77~km link used in these tests, the 10 minute phase drift never exceeded 0.56~rad and all drifts within one standard deviation of the mean had a magnitude less than 0.08~rad. However, the 10 minute phase drift is dependent on both systematic and random phase fluctuations, so it is not known how the magnitude of the phase drifts will scale with increasing link length. Also, unlike for the Allan deviation results, it was not possible to determine, from the data obtained, how much of the observed phase drift was due to the stabilized frequency transfer system, and how much can be attributed to other systems, such as the E/O systems. For this reason, bench tests in the laboratory are a better method for validating the stabilized transfer system against this SKA systems requirement. Despite this, these results are a strong indication that this stabilized frequency transfer system is capable of exceeding the 10-minute phase drift requirement.

These tests have also highlighted potential problems with the use of multiple synthesizers to provide phase coherent signals to separate antennas, even when the reference frequencies to those synthesizers have been successfully phase-stabilized. Care must be taken to ensure that alternative stabilized frequency reference transfer systems being considered for the SKA do not suffer from phase drift to the extent that it is a detriment to the performance of the telescope. Synthesizer phase-drift rates such as those seen in these tests would reduce the sensitivity of the array and, under some conditions, limit the imaging capability.

\acknowledgments

This paper describes work being carried out for the SKA Signal and Data Transport (SaDT) consortium as part of the Square Kilometre Array (SKA) project. The SKA project is an international effort to build the world's largest radio telescope, led by the SKA Organisation with the support of 10 member countries.   
      The Australia Telescope is funded by the Commonwealth of Australia for operation as a National Facility managed by CSIRO.
      We are very grateful for the support provided by the staff at CSIRO Astronomy and Space Science and the Australia Telescope National Facility.
      Thank you to Simon Stobie and Gavin Siow for their contributions to the construction and testing of the frequency transfer system.
      This work was supported by funds from the University of Manchester and the University of Western Australia.

\end{document}